# Charge-Density-Wave Phase Transitions in Monolayer 1*T*-$TaS_2$ from Universal Machine Learning Molecular Dynamics

Valentina Nesterova,[1] Tribhuwan Pandey[2], Tom Berlijn[3,4], Fariborz Kargar,[1] Lucas Lindsay[3*], Konstantin Klyukin[1*]

[1] Department of Mechanical and Materials Engineering, Auburn University, Auburn, Alabama 36849, United States
[2] Department of Physics, University of Antwerp, Groenenborgerlaan 171, Antwerp B-2020, Belgium
[3] Materials Science and Technology Division, Oak Ridge National Laboratory, Oak Ridge, Tennessee 37831, United States
[4] Center for Nanophase Materials Sciences, Oak Ridge National Laboratory, Oak Ridge, Tennessee 37831, United States

**ABSTRACT**. Charge-density-wave (CDW) phases in 1*T* transition-metal dichalcogenides arise from strong electron-phonon coupling and accompanying lattice instabilities. Capturing their temperature-dependent structural evolution using conventional first-principles molecular dynamics (MD) remains challenging because of the large supercells and extensive finite-temperature sampling required. Here, we combine density functional theory (DFT), universal machine-learning interatomic potentials (MLIPs), MD, and temperature-dependent effective potential phonon calculations to investigate the structural and vibrational signatures of CDW transitions in monolayer 1*T*-$TaS_2$. Benchmarking against DFT displacement energies identifies UMA-s-1p1 universal machine learning potentials with sufficient accuracy for subsequent finite-temperature simulations. Our results show that large-scale MD simulations reproduce the experimentally observed phase transition sequence from the low-temperature Star-of-David (SoD) distorted structure to the high-temperature primitive hexagonal structure, as quantified by the number of Ta atoms attributed to SoDs. Heating–cooling cycles exhibit thermal hysteresis, and upon cooling, the system freezes into a multi-domain state in which α and β CDW chiralities nucleate independently and persist to the lowest temperatures. These findings demonstrate that carefully benchmarked universal MLIPs can provide a scalable framework for finite-temperature studies of CDW materials.

## I. INTRODUCTION

Bulk and two-dimensional transition metal dichalcogenides (TMDCs) have attracted significant attention because of their tunable bandgaps, strong spin-orbit and electron–phonon couplings, rich interplay between competing phases and lattice instabilities, and charge density wave (CDW) formation. [1–5] Among these materials, the 1*T* phase of $TaS_2$ has attracted particular interest because it exhibits multiple CDW phases, including a technologically relevant phase transition near room temperature (RT) that can be triggered by thermal, electrical, and optical stimuli. [1,5] This near-room-temperature transition is accompanied by substantial changes in electronic properties and has enabled a variety of device concepts, including oscillators, memristive devices, and neuromorphic computing elements. [2,6,7] At high temperatures ($T > 550$ K), 1*T*-$TaS_2$ is in the normal metallic (NM) state with a hexagonal primitive crystal structure. Upon cooling, it undergoes a sequence of CDW transitions, first to an incommensurate CDW (ICDW) phase ($350 \text{ K} < T < 550$ K), followed by the nearly incommensurate CDW (NCCDW) phase ($180 \text{ K} < T < 350$ K) consisting of commensurate "Star-of-David" (SoD) domains embedded within the incommensurate primitive lattice. [8–12] Below 180 K, the system transforms into a commensurate CDW (CCDW) phase characterized by a fully developed SoD distortion throughout the crystal. [3,13–15] These phase transitions in 1*T*-$TaS_2$ give rise to distinct transport and vibrational signatures that have been extensively characterized experimentally in 1*T*-$TaS_2$ and are broadly observed in other CDW TMDCs. [3,4,8,13,15–19]

Despite these advances, a complete microscopic understanding of the temperature-dependent structural evolution and lattice dynamics of 1*T*-$TaS_2$ remains unclear. A major challenge is describing the intermediate NCCDW phase, whose coexistence of commensurate SoD domains and incommensurate regions spans length scales that are inaccessible to conventional first-principles simulations. As a result, previous DFT studies have primarily been limited to the static, ground-state (0 K) calculations of the commensurate phase [12] or analyses of lattice instabilities in the primitive structure [17], leaving the finite-temperature evolution of the intermediate CDW phases largely unexplored. Yet, finite-temperature effects are essential for understanding the structural evolution, phase stability, and vibrational properties across the CDW transitions.

Recent works on the related compound 2D $NbSe_2$ have demonstrated that machine learning interatomic potentials (MLIPs) can capture the subtle structural and vibrational signatures of CDW phases, including transition temperatures and phonon renormalization, [20,21] at a fraction of the cost of direct *ab initio* molecular dynamics. However, training system-specific MLIPs requires carefully curated datasets and extensive hyperparameter tuning, limiting their transferability to new materials or phases. Universal machine learning interatomic potentials (uMLIPs), trained on broad materials databases, offer a compelling alternative: they require no material-specific training and can be deployed directly across the TMDC family for systematic finite-temperature studies of CDW systems such as 1$T$-$TaS_2$.

*Corresponding authors: lindsaylr@ornl.gov and klyukin@auburn.edu

In this work, we employ a universal machine-learning potential-based molecular dynamics coupled with a temperature-dependent effective potential (TDEP) approach to systematically study the temperature-dependent structural evolution and phonon spectra of monolayer 1$T$-$TaS_2$. Without any material-specific training, the universal potential enables large-scale finite-temperature simulations that naturally capture the intermediate NCCDW phase, a long-standing challenge for conventional first-principles simulations. Our simulations reproduce the experimentally observed sequence of CDW phase transitions and the accompanying temperature-dependent structural evolution in 1$T$-$TaS_2$. The computed structural and vibrational properties provide microscopic insights into the mechanisms governing the evolution, stabilization, and lattice dynamics of the CDW phases. These findings complement earlier experimental studies, [3,8,9] bridging large-scale finite-temperature simulations and experimental observations and contributing to a deeper understanding of the interplay between lattice dynamics and electronic properties in CDW 2D TMDCs.

## II. METHODS

### A. Density Functional Theory (DFT).

First-principles calculations were performed using the Vienna Ab initio Simulation Package (VASP) [22–24] with the projector augmented-wave (PAW) method [25] and the Perdew-Burke-Ernzerhof (PBE) generalized gradient approximation. [26] The monolayer geometry was achieved by creating a vacuum gap of 12 Å between 1$T$-$TaS_2$ layers. To assess the role of electronic correlations, selected calculations were repeated with an on-site Hubbard U correction (DFT+U) using the Dudarev formulation [27] with U = 2.5 eV applied to the Ta 5$d$ states (based on literature data [28,29]) and including spin polarization in a ferromagnetic (FM) configuration. These calculations were compared to those without U and spin polarization (non-magnetic). [28,29] For spin-polarized calculations of the CCDW phase, the FM structure was found to be more energetically favorable than the non-magnetic and two other antiferromagnetic configurations considered. For the NM phase, the ferromagnetic state also was more favorable than the non-magnetic state. The plane-wave cutoff energy was set to 500 eV and Brillouin zone sampling was performed on a 4×4×1 $k$-points mesh. Structural relaxations were performed with an atomic force threshold 0.01 eV/Å, $a$ and $b$ cell parameters were relaxed, while the $c$ cell parameter was fixed. Phonon dispersions at the harmonic level were calculated with the Phonopy package [30] using the finite-difference approach with cell sizes varying from 3×3×1 (27 atoms) to 2√13×2√13×1 (156 atoms) and with 4×4×1 Gamma-centered $k$-point meshes.

### B. Universal Machine Learning Interatomic Potentials (uMLIPs).

Two sets of universal machine learning interatomic potentials were evaluated: MACE-MP-0 [31] and the UMA s-1p1 [32] trained on the OMat-24 dataset. [33] Both uMLIPs were benchmarked here against DFT PBE with both U = 0 and U = 2.5 eV, and the relative energies were computed on a dataset of structures generated by applying random displacements in the range 0.01-0.15 Å to both the high temperature NM and the low temperature CCDW 1$T$-$TaS_2$ monolayer structures with 2√13 × 2√13 supercells. The UMA s-1p1 model was selected for all subsequent molecular dynamics simulations based on its overall accuracy.

### C. uMLIP-based Molecular Dynamics (MD) and Structural Analysis.

MD simulations were carried out using the UMA s-1p1 potential on supercells based on hexagonal primitive cell of 1$T$-$TaS_2$ monolayer. We performed our calculations on large supercells containing 2028 atoms, as discussed later in the results and discussion section. Each simulation was equilibrated at the target temperature for 50-100 ps in the NVT Langevin ensemble [34] with a timestep of 1 fs. Cooling and heating simulations used to demonstrate hysteresis behavior were performed as a series of consecutive MD calculations for these 2028 atom cells. For cooling, the initial temperature was set to 500 K, for heating – 60 K. Each subsequent temperature was initialized from the final frame of the preceding simulation, using both positions and velocities as a starting configuration. At each temperature, the system was equilibrated for 50 ps. All MD calculations were carried out using the atomistic simulation python environment (ASE). [35]

Star of David (SoD) patterns were identified during the analysis of MD trajectories of the large 2028 atom supercells, using Ta–Ta distances from the DFT-relaxed CCDW cell as the reference for the ideal SoD geometry. At each simulated temperature, the Ta atomic positions were averaged over the last 15 ps of simulation to obtain a representative averaged distribution of atoms. Each Ta atom was then evaluated as a potential SoD center based on its six nearest Ta neighbors (1NN) and its second coordination shell (2NN), defined within a radius of 5.8 Å. A candidate center was retained when the distances to all six 1NN Ta atoms were below 3.23 Å, including a temperature-dependent tolerance for thermal fluctuations. The candidate was a SoD center when two additional criteria were satisfied: *(i)* its second coordination shell (2NN), defined within a radius of 5.8 Å, contained at least three Ta atoms, and *(ii)* at least three of its 1NN Ta atoms exhibited the characteristic SoD bonding motif, consisting of three short Ta–Ta bonds (<3.23 Å) and one long Ta–Ta bond (≥3.7 Å). Requiring at least three, rather than all six, 2NN atoms to satisfy this criterion allowed the method to identify partially formed SoD clusters at grain boundaries, where the ideal bonding pattern is disrupted by the local mismatch between neighboring CDW domains. All Ta atoms satisfying these criteria were assigned to the corresponding SoD cluster. The distance tolerances used to define the first- and second-neighbor shells were set to the mean Ta root-mean-square displacement (RMSD) computed over the trajectory window, allowing the classification criteria to adapt to the temperature-dependent thermal vibrations. The scripts for this analysis are provided on Figshare. [36] The same procedure was applied to atomic positions averaged over 2 ps windows to visualize the time evolution of SoD ordering within a fixed temperature. The resulting trajectories are provided as video files on Figshare storage. [36]

### D. Temperature-Dependent Effective Potential (TDEP) for Phonons.

Temperature-dependent phonon dispersions were extracted from MD trajectories using the TDEP method. [37,38] The last 15 ps were used for the 1*T*-$TaS_2$ system with 2028 atom supercells, every $10^{th}$ snapshot was used. Effective interatomic force constants were extracted by fitting to the forces sampled during MD, and the resulting dynamical matrices were used to obtain phonon dispersions of the NM cell structure. A cutoff radius of 8 Å used; rotational and Huang invariances were not enforced. Lattice thermal expansion was not accounted for in the MD simulations and TDEP phonon calculations. A script for converting initial UMA-generated files into input files for the TDEP software is provided at Figshare storage. [36]

## III. RESULTS AND DISCUSSION

Large supercell simulations are required to investigate the phase transitions in monolayer 1*T*-$TaS_2$, which are computationally prohibitive for routine first-principles calculations, but easily accessible for universal machine-learning interatomic potentials (uMLIPs). However, many existing uMLIPs, including those considered here, are trained primarily on large DFT-based datasets such as the Materials Project, [39] Alexandria, [40,41] OMat-24, [33] which are often generated without Hubbard corrections or spin-polarized magnetic moments. Therefore, it is important to first establish how sensitive the electronic structure and phase transitions in monolayer 1*T*-$TaS_2$ are to the choice of computational parameters before benchmarking existing uMLIPs against first-principles calculations to validate their use in large-scale simulations of CDW phase transitions.

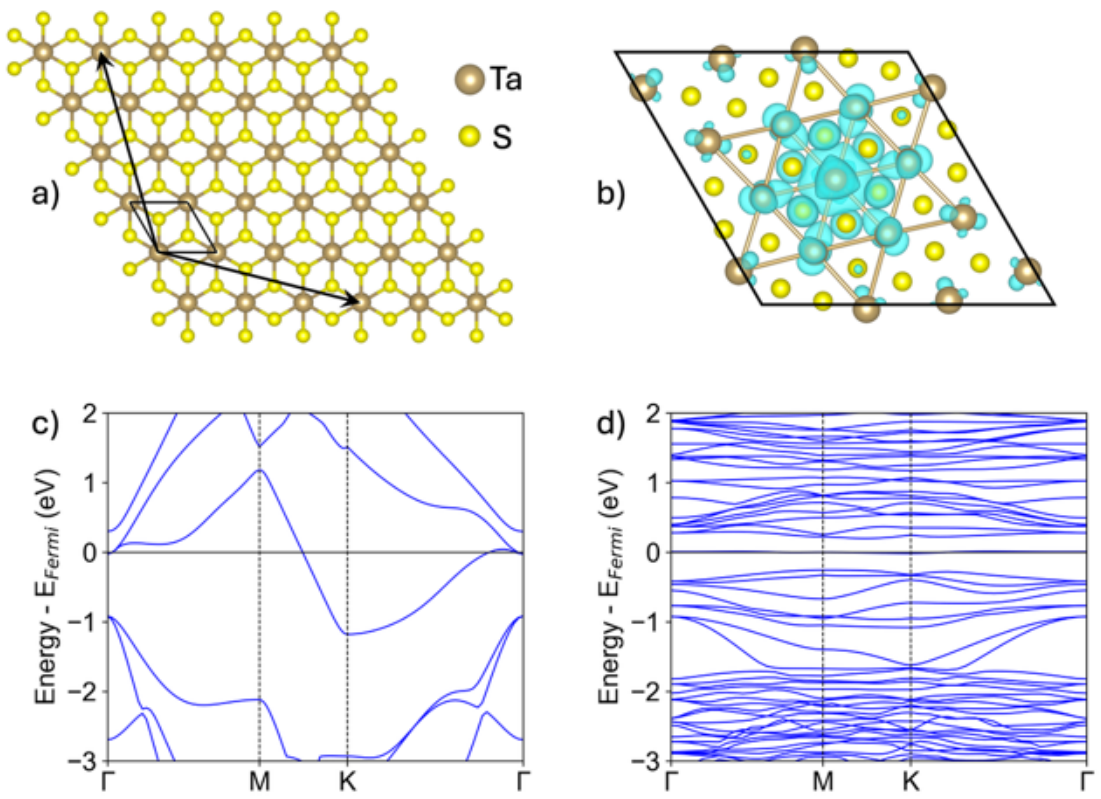

FIG. 1 (a) NM and (b) CCDW unit cells of 1*T*-$TaS_2$ monolayer, along with their corresponding electronic band structures shown in (c) and (d), respectively. The vectors in panel (a) indicate the transformation from the NM cell to the CCDW supercell. Panel (b) also shows the isosurface of the highest occupied state, plotted at an isovalue of 0.0015 a.u., illustrating electronic localization in the CCDW phase. Electronic structures were calculated using non-spin-polarized PBE with U = 0.

We begin by comparing the undistorted hexagonal 1*T*-$TaS_2$ structure, which approximates the high-temperature state associated with the NM phase, with the low-temperature CCDW phase, shown in **Figures 1a** and **1b**, respectively, using several DFT settings. **Figures 1c** and **1d** show the electronic band structures of these phases computed with non-spin-polarized PBE and U = 0. The NM phase exhibits a metallic band structure, with Ta-derived bands crossing the Fermi level, consistent with the high-temperature metallic character reported previously. [13,14] In contrast, formation of the SoD distortion in the CCDW phase leads to strong electronic localization within the cluster of 13 Ta atoms (**Figure 1b**). As a result, the CCDW band structure has a pronounced flat band near the Fermi level, originating primarily from the nearly isolated Ta orbital at the center of each SoD cluster. The unfolded band structure of the CCDW phase, projected onto the primitive NM phase Brillouin zone, is provided in **Figure S1**. Comparing these results with calculations performed using spin polarization and U = 2.5 eV (**Figure S2**), we find that the main qualitative features are preserved, *i.e.*, the NM phase and the electron-localized CCDW phase with a flat band near the Fermi level are reproduced regardless of the spin polarization and Hubbard corrections.

Besides the electronic structure, we also examine the sensitivity of the phase transition energetics to spin polarization and the Hubbard correction (**Figure S3**). The transition enthalpy between the two phases is weakly affected by their choices. We obtain ΔH = 0.028 eV per f.u. for U = 0 and ΔH = 0.019 eV per f.u. for U = 2.5 eV. The slightly larger transition enthalpy obtained at U = 0 may indicate a modest overestimation of the CCDW phase stability, since a higher thermal energy would be required to destabilize the CCDW state. These findings establish that non-spin-polarized PBE with U = 0 captures the relative structural energetics and the main qualitative electronic features relevant to the present structural analysis, thus establishing confidence that the uMLIPs trained without these parameters may provide reasonable description of the lattice dynamics of monolayer 1*T*-$TaS_2$.

Next, we benchmark two models, UMA-s-1p1-omat [32] and MACE-MP-0, [31] against PBE-calculated displacement energies (non-spin-polarized; U = 0) for the NM and CCDW phases (**Figure 2**). Atomic displacements ranging from 0.01 to 0.15 Å, that correspond to the temperature fluctuations near phase transition region (**Figure S4**), are used to sample the local energy landscape associated with thermally accessible vibrational distortions. MACE-MP (**Figure 2a**) systematically underestimates displacement energies. In contrast, UMA s-1p1 (**Figure 2b**) shows excellent agreement with PBE energies across the full displacement range. UMA also reproduces the relative stability of the two phases, yielding a transition enthalpy of ΔH=0.0272 eV per f.u., in close agreement with the DFT value. The uncertainty grows with displacement amplitude for both models, as expected, but remains reasonable for UMA s-1p1 with mean absolute error (MAE) of 0.048 eV per f.u. Comparable performance is obtained when the benchmark is repeated against DFT+U data with U = 2.5 eV and spin polarization (**Figure S5**), confirming that phase transformations in monolayer 1*T*-$TaS_2$ are not strongly sensitive to correlations and magnetism as described by these DFT settings. On the basis of this benchmarking, all subsequent MD simulations employ the UMA s-1p1 model.

Having established that the UMA-s-1p1 model accurately reproduces DFT energies for both the NM and CCDW phases, and that calculations are relatively insensitive to correlations and magnetic structure, we now employ this MLIP in MD simulations to investigate the finite-temperature structural behavior of monolayer 1*T*-$TaS_2$ and resolve the phase transition to the NCCDW phase. To capture the long-range ordering of CDW chiral domains, we selected a large supercell containing 2028 atoms. The CCDW superlattice can adopt two mirror-related orientations, corresponding to the energetically degenerate α and β chiral variants. [42,43] The simulation cell was designed to accommodate both variants simultaneously, allowing domains of each chirality to nucleate independently. (**Figure S6**) Starting from the fully commensurate CCDW phase at low temperature, we conducted sequential heating to observe the progressive structural transformations from CCDW to NCCDW, further to ICDW-like phase, and to the NM hexagonal phase of 1*T*-$TaS_2$.

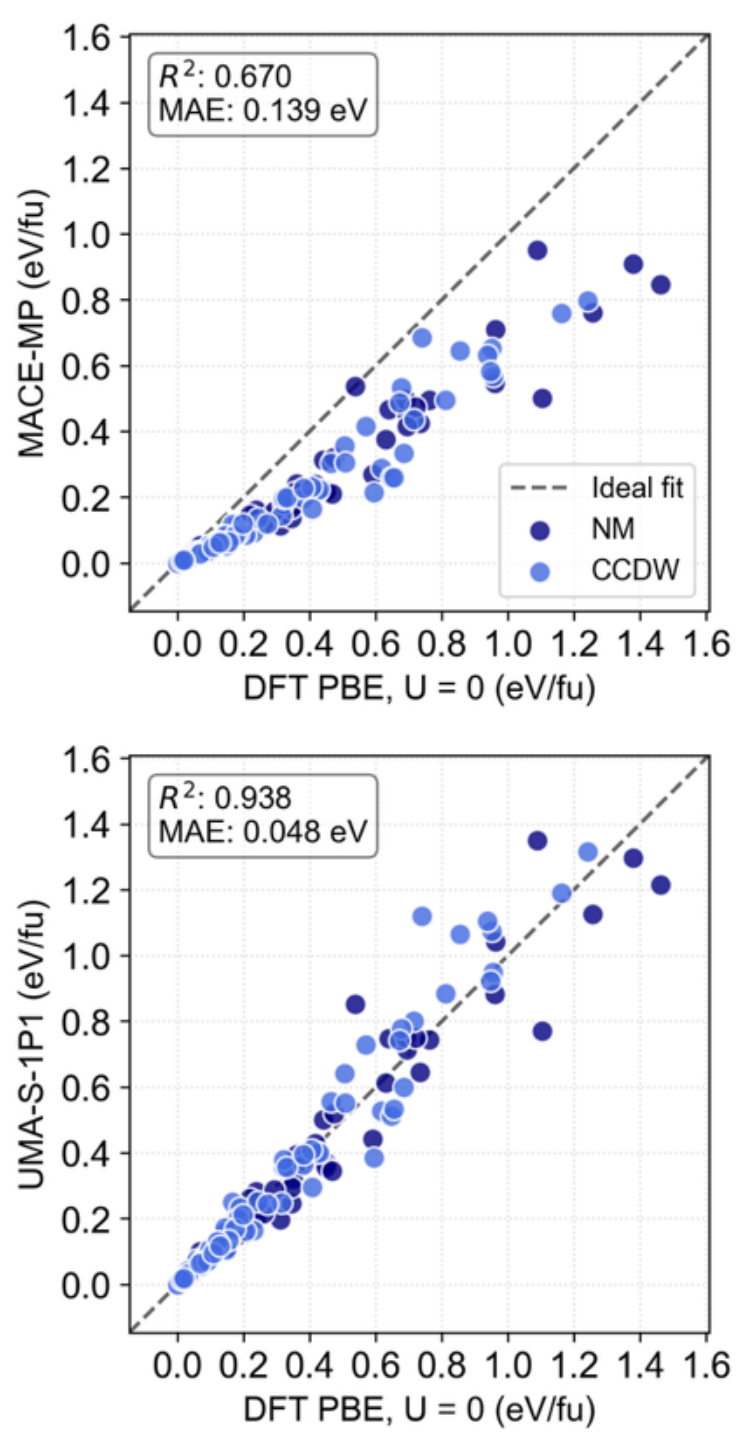

FIG. 2 Benchmarking (a) MACE-MP-0 and (b) UMA s-1p1 OMat-24 models against DFT PBE energies. The benchmarking dataset was generated by applying random atomic displacements ranging from 0.01 to 0.15 Å to both NM and CCDW 1*T*-$TaS_2$ structures.

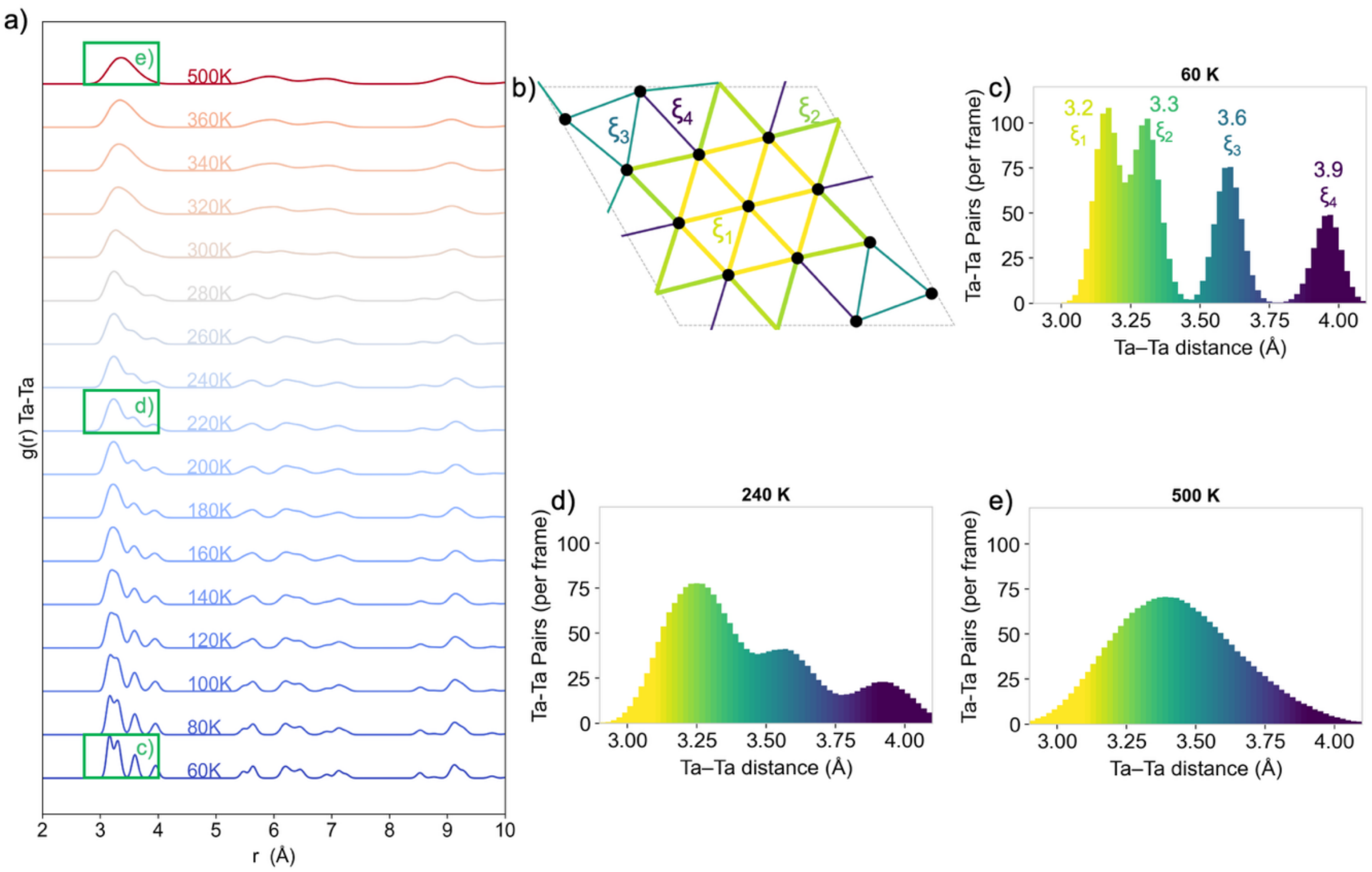

FIG. 3 . Temperature evolution of Ta–Ta bond distributions in monolayer 1*T*-$TaS_2$ during heating, calculated using UMA s-1p1 MD for 2028-atom supercells. (a) Ta–Ta pair correlation functions at selected temperatures. (b) Schematic representation of four symmetry-inequivalent Ta–Ta bonds (labeled $\xi_1$-$\xi_4$) within a cell containing a single SoD pattern, only Ta atoms are shown. (c–e) Nearest-neighbor Ta–Ta bond-length distributions at 60, 240, and 500 K.

To characterize the structural order of monolayer 1*T*-$TaS_2$, we examine the Ta-Ta pair correlation function g(r) (**Figure 3a**), which provides a fingerprint for each phase and can be used to track the phase transitions. At low temperatures, the 1*T*-$TaS_2$ monolayer displays a characteristic splitting of the nearest-neighbor Ta–Ta peak into four components, corresponding to the four symmetry-inequivalent bond lengths ($\xi_1$–$\xi_4$) within the CCDW (**Figure 3b,c**). This splitting is a direct consequence of the Star-of-David distortion: as the surrounding Ta atoms contract toward the center of each cluster, the single nearest-neighbor distance of the undistorted lattice is replaced by a set of shorter intra-cluster bonds and longer inter-cluster separations. With increasing temperature, thermal motion progressively broadens (**Figure 3d**) and merges these components until, in the high-temperature phase, g(r) recovers a single sharp peak at 3.4 Å reflecting the uniform nearest-neighbor distance of the undistorted NM phase (**Figure 3e).**

While the splitting of the Ta–Ta pair correlation function peak qualitatively captures the phase transition in the 200-300 K region (**Figure 3a**), this metric cannot quantify the fraction of Ta atoms retaining local SoD order, nor resolve the coexistence of ordered and disordered regions characteristic of the mixed NCCDW phase. To address this limitation, we introduce a structural metric based on identifying individual SoD clusters: for each Ta atom, we determine whether it belongs to a SoD pattern based on characteristic nearest-neighbor distances (see Methods and Figshare files for details [36]). This metric provides a quantitative measure of SoD ordering as a function of temperature, ranging from 1 in the CCDW phase to 0 in the NM phase.

To track the evolution of CDW order across the phase transitions, we performed sequential heating and cooling simulations. Upon heating, the SoD-counting metric reveals a sharp CCDW-to-NCCDW transition at ~290 K (**Figure 4a,c**). At temperatures just above 300 K, we observe NCCDW-like states characterized by coexistence of well-defined SoD-ordered and disordered regions, consistent with the nearly commensurate mixed phase reported experimentally for 1*T*-$TaS_2$ in this temperature range. With further heating, the system transitions into an ICDW-like state in which SoD clusters form and dissolve continuously,

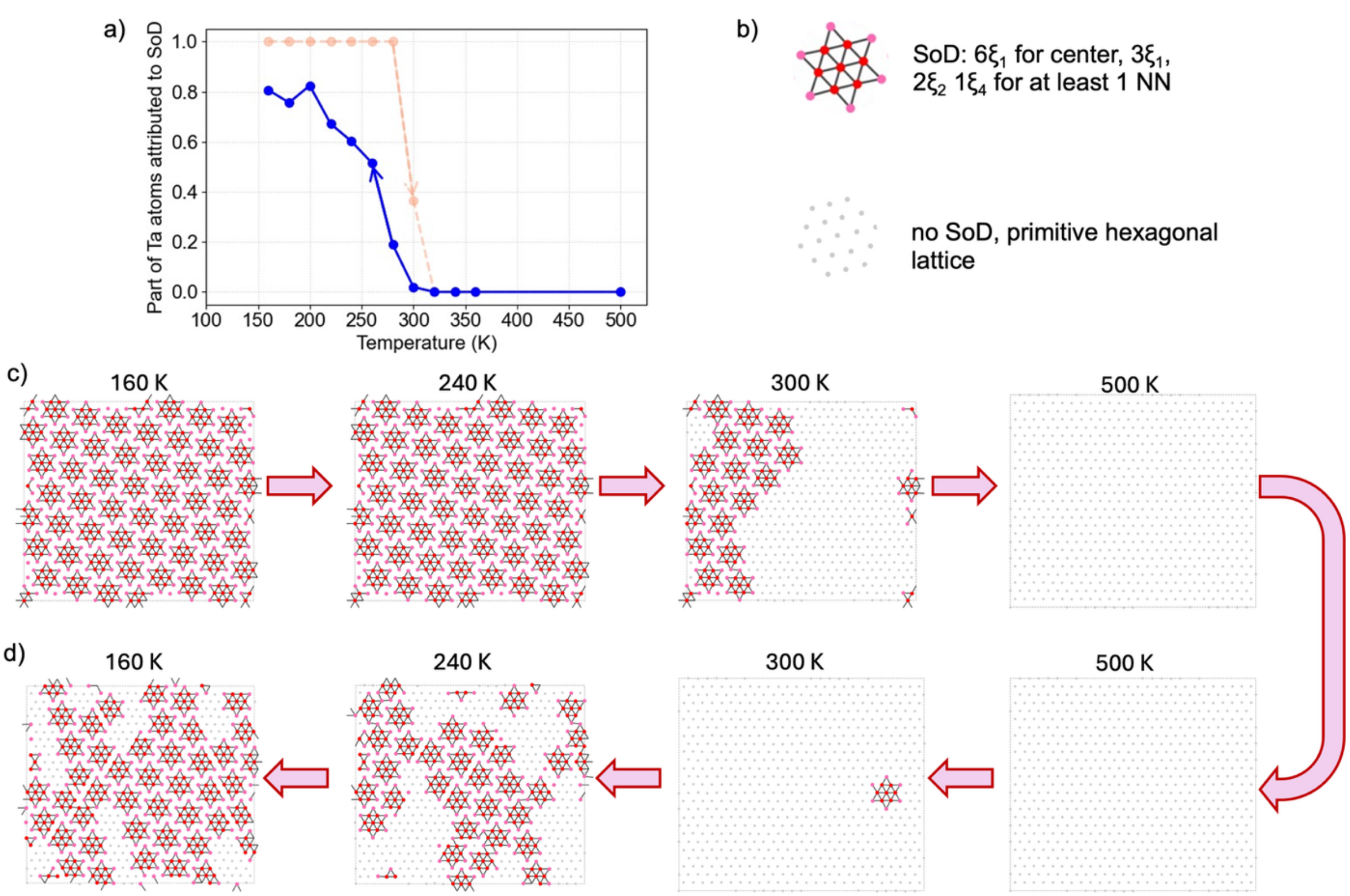

FIG. 4 (a) Temperature dependence of the part of Ta atoms in the cell attributed to SoD from MD simulations, showing a phase transition in the 290–300 K range, pink line shows heating regime, while blue line shows cooling regime; (b) Schematic legend distinguishing Star-of-David (SoD) clusters and primitive hexagonal patterns observed in the MD trajectories; (c) and (d) Temperature evolution of the simulated structure during the heating/cooling cycle, showing a mixed state at intermediate temperatures. Only Ta atoms are shown; each image is averaged over 15 ps. The simulation cell contains 2028 atoms in total.

appearing only transiently at random positions so that no long-range-periodic arrangement is maintained. By ~500 K, the system adopts a fully hexagonal metallic structure with no SoD appearing (**Figure 4c**).

Subsequent cooling from 500 K reveals a phase transition at approximately 260 K (**Figure 4a,d**), roughly 30 K below the heating transition, producing a thermal hysteresis. Notably, cooling does not recover a single fully ordered CCDW phase. Instead, the structure fragments into multiple CCDW domains separated by boundaries where SoD patterns are incomplete or absent (**Figure 4d**). Individual snapshots show that these boundaries separate domains of opposite CDW chirality (α and β), reflecting independent nucleation of chiral variants at different sites (see **Figure S7**), as previously reported in the literature. [42,43] At the cooling conditions considered here, the multi-domain structure persists to the lowest temperatures rather than converting into fully ordered CCDW structure, consistent with previous reports of a suppressed CCDW phase in monolayer 1*T*-$TaS_2$. [44] The interlayer coupling [44,45] and defect-mediated pinning [13,46] may substantially alter domain nucleation and phase stability in thin 1*T*-$TaS_2$ films, but these effects are beyond the scope of the present study.

The Ta–Ta bond-length distributions provide additional insight into the local structure of the multidomain state. Within the CCDW domains, distinct peaks appear near 3.2, 3.3, 3.6, and 3.9 Å, reflecting the characteristic Ta–Ta separations within the Star-of-David clusters. In the regions separating these domains, the NCCDW-like state near the transition temperature is dominated by Ta–Ta bonds of approximately 3.4 Å, close to the spacing of the undistorted hexagonal lattice (**Figure S8**). At lower temperatures, however, the interdomain bond-length distribution shifts toward approximately 3.2 Å. As the domain-wall regions become narrower and more strongly confined between neighboring CCDW domains, they have insufficient space to fully relax toward the undistorted geometry of NM hexagonal lattice, resulting in shorter Ta–Ta separations.

To further characterize the vibrational signatures of the transition, **Figure 5** presents the temperature-dependent phonon dispersions of monolayer 1*T*-$TaS_2$ calculated for the NM cell structure from MD trajectories of the heating cycle using the TDEP approach. [37] At low temperatures (blue curves), several phonon branches exhibit pronounced softening and become imaginary (shown as negative values in **Figure 5**), indicating the dynamical instability of the NM phase toward CDW distortion. Above the transition temperature, the dispersion resembles that of the conventional metallic phase, with no unstable modes remaining in the primitive Brillouin zone. The calculated dispersions are in good agreement with harmonic DFT phonon dispersions computed on a dense $q$-mesh with large supercells (**Figure S9**). The soft branch in the Γ-M region corresponds to Ta vibrations in $10\bar{1}0$-$\bar{1}010$ direction, and modes in the K-Γ region correspond to Ta vibrations in $\bar{1}100$-$1\bar{1}00$ and $11\bar{2}0$-$\bar{1}\bar{1}20$ directions (**Figure S10**). These modes disappear in the range of temperatures from 100 K to 300K as the imaginary frequencies shift to positive values (red curves), indicating dynamical stabilization of the NM phase by anharmonic thermal fluctuations. This behavior is consistent with the mixed structural state observed in **Figure 4c.**

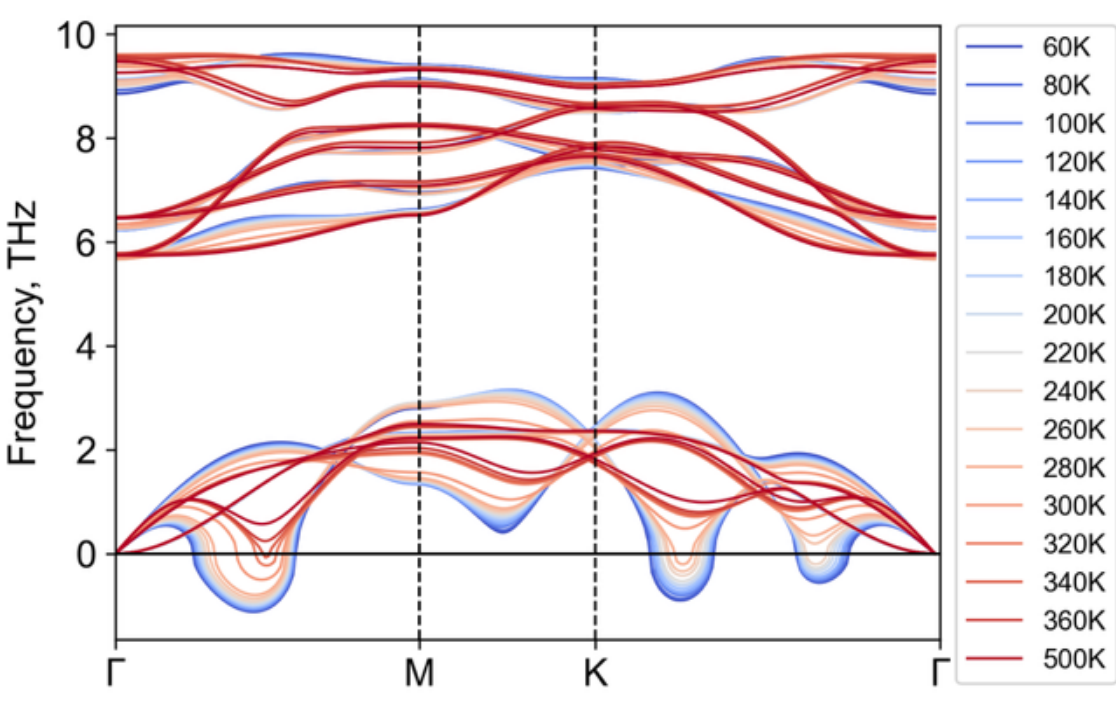

FIG. 5 Phonon dispersions of monolayer $1T$-$TaS_2$ determined for the NM structure as a function of temperature, with curves colored from blue at low temperature to red at high temperature. The interatomic force constants (IFCs) were calculated from molecular dynamics trajectories generated through the subsequent heating using the UMA s-1p1 model for a simulation cell containing 2028 atoms.

Several limitations and opportunities for further improvement are noted here. First, because the simulations use uMLIPs without material-specific fine-tuning, quantitative transition temperatures across a broad range of CDW materials should be interpreted with caution, especially for systems whose energetics are sensitive to Hubbard corrections, magnetism, or other electronic-structure effects. Second, limitations associated with finite cutoff radii and the limited ability of current MLIPs to capture interlayer coupling may affect the resolution of long-wavelength incommensurate or nearly commensurate domain structures. Third, spin–orbit coupling and magnetic interactions may play important roles in some CDW systems, but they are typically not included explicitly in uMLIP-based dynamics. Despite these limitations, the agreement among DFT benchmarking, structural order metrics, and temperature-dependent phonon trends suggests that this approach captures the dominant lattice contributions to the CDW transition in monolayer $1T$-$TaS_2$.

## IV. CONCLUSIONS

We have demonstrated that a pretrained universal machine-learning interatomic potential, benchmarked against DFT but requiring no material-specific training, can resolve key finite-temperature lattice signatures associated with CDW ordering in monolayer of $TaS_2$. Benchmarking against DFT displacement energies, including tests of sensitivity to Hubbard corrections and spin polarization, identified UMA-s-1p1 as a reliable model for the relevant lattice degrees of freedom. UMA-based MD simulations of 2028-atom supercells reveal a sharp loss of SoD order near room temperature during heating, quantified by the number of Ta atoms attributed to SoD patterns. Across the transition, the system passes through an NCCDW-like state in which SoD-ordered and primitive regions coexist, followed by an ICDW-like regime of transient, randomly nucleating clusters, before recovering the undistorted metallic structure at high temperatures. Heating and cooling cycles produce a thermal hysteresis. Cooling the system locks into a multi-domain configuration in which α and β CDW chiralities nucleate independently and persist to the lowest temperatures, reproducing the formation of multidomain structures reported experimentally. TDEP phonon calculations show the progressive stabilization of CDW-related soft modes with increasing temperature, providing a vibrational signature of anharmonic stabilization of the high-temperature phase.

Together, these results show that carefully validated universal interatomic potentials can provide a scalable framework for investigating finite-temperature lattice transformations in CDW materials. The framework opens the door to systematic studies of domain dynamics, finite-size and strain effects, defect-mediated pinning and its impact on CDW transformations, as well as to investigations of multilayer systems in which interlayer coupling can further influence phase stability and domain evolution.

## ACKNOWLEDGMENTS

Initial calculations and analysis (V.N., T.B., L.L.) were supported by the U.S. Department of Energy, Office of Science, Basic Energy Sciences, Materials Sciences and Engineering Division. V.N. and K.K. acknowledge funding support from the National Science Foundation through DMREF-2324156. This research used resources of the National Energy Research Scientific Computing Center (NERSC), a Department of Energy Office of Science User Facility using NERSC award BESERCAP-m1057. This work utilized Stampede3 at Texas Advanced Computing Center (TACC) at The University of Texas at Austin through allocation MAT250073 from the Advanced Cyberinfrastructure Coordination Ecosystem: Services & Support (ACCESS) program, [47] which is supported by U.S. National Science Foundation grants #2138259, #2138286, #2138307, #2137603, and #2138296.

**SUPPORTING INFORMATION**

# Temperature-Driven Phase Transitions in Charge-Density-Wave 1*T*-$TaS_2$ Material from Machine Learning Molecular Dynamics

Valentina Nesterova,[1] Tribhuwan Pandey[2], Tom Berlijn[3,4], Fariborz Kargar,[1] Lucas Lindsay[3*], Konstantin Klyukin[1*]

1- Department of Mechanical and Materials Engineering, Auburn University, Auburn, Alabama 36849, United States
2- Department of Physics, University of Antwerp, Groenenborgerlaan 171, Antwerp B-2020, Belgium
3- Materials Science and Technology Division, Oak Ridge National Laboratory, Oak Ridge, TN 37831, United States
4- Center for Nanophase Materials Sciences, Oak Ridge National Laboratory, Oak Ridge, Tennessee 37831, USA

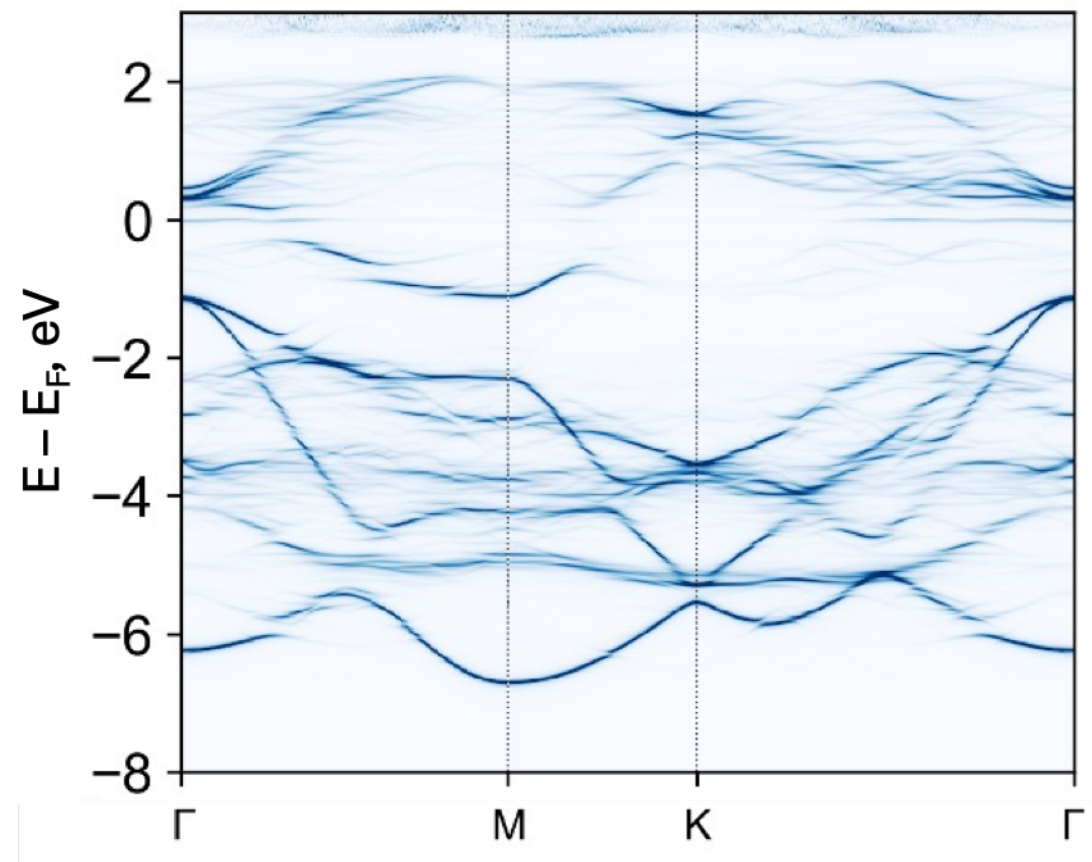

**Figure S1**. CDW electronic structure of 1*T*-$TaS_2$ unfolded to the NM hexagonal cell.

*Corresponding authors: lindsaylr@ornl.gov and klyukin@auburn.edu

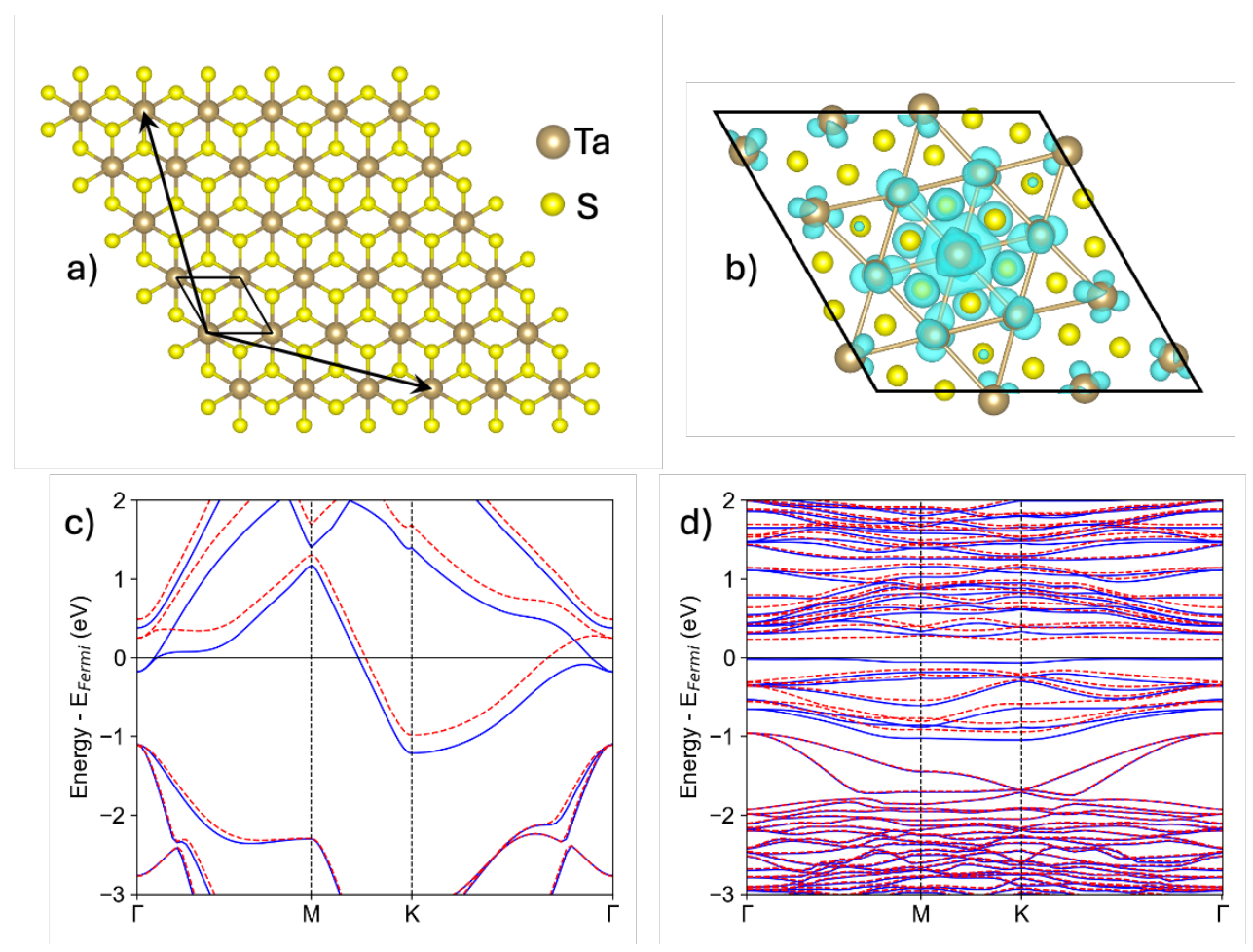

**Figure S2**. (a) NM and (b) CCDW primitive cells of monolayer 1*T*-$TaS_2$ with their electronic band structures, (c) and (d), respectively, red and blue shows spin up and spin down channels, respectively. Vectors in (a) represent transition from the NM cell to the CCDW cell. Figure (b) includes electronic localization representation, isosurface level = 0.0015. Electronic structures were calculated with spin-polarized PBE, U = 2.5.

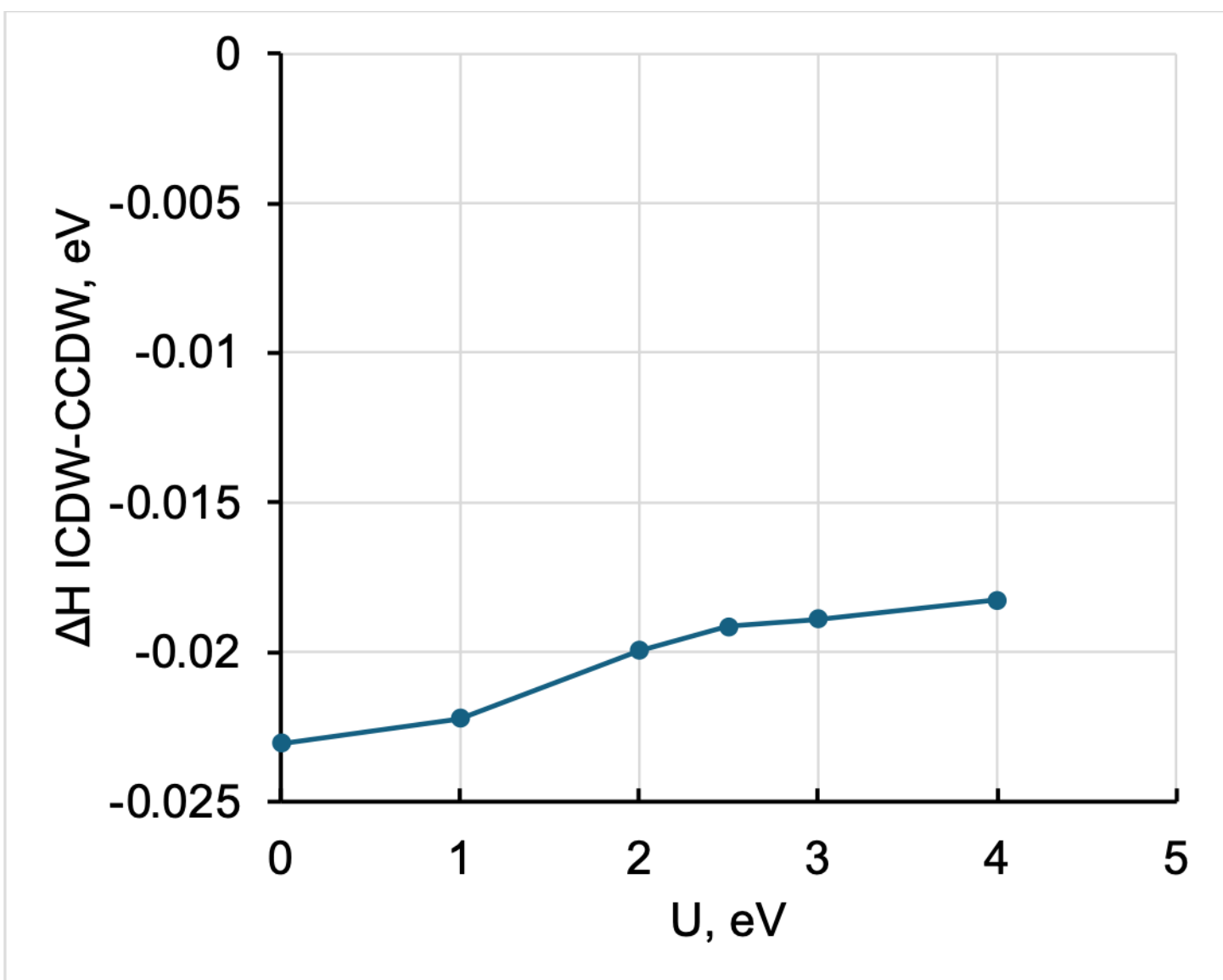

**Figure S3**. Enthalpy of NM/CCDW phase transition across different Hubbard U corrections, applied to Ta atoms.

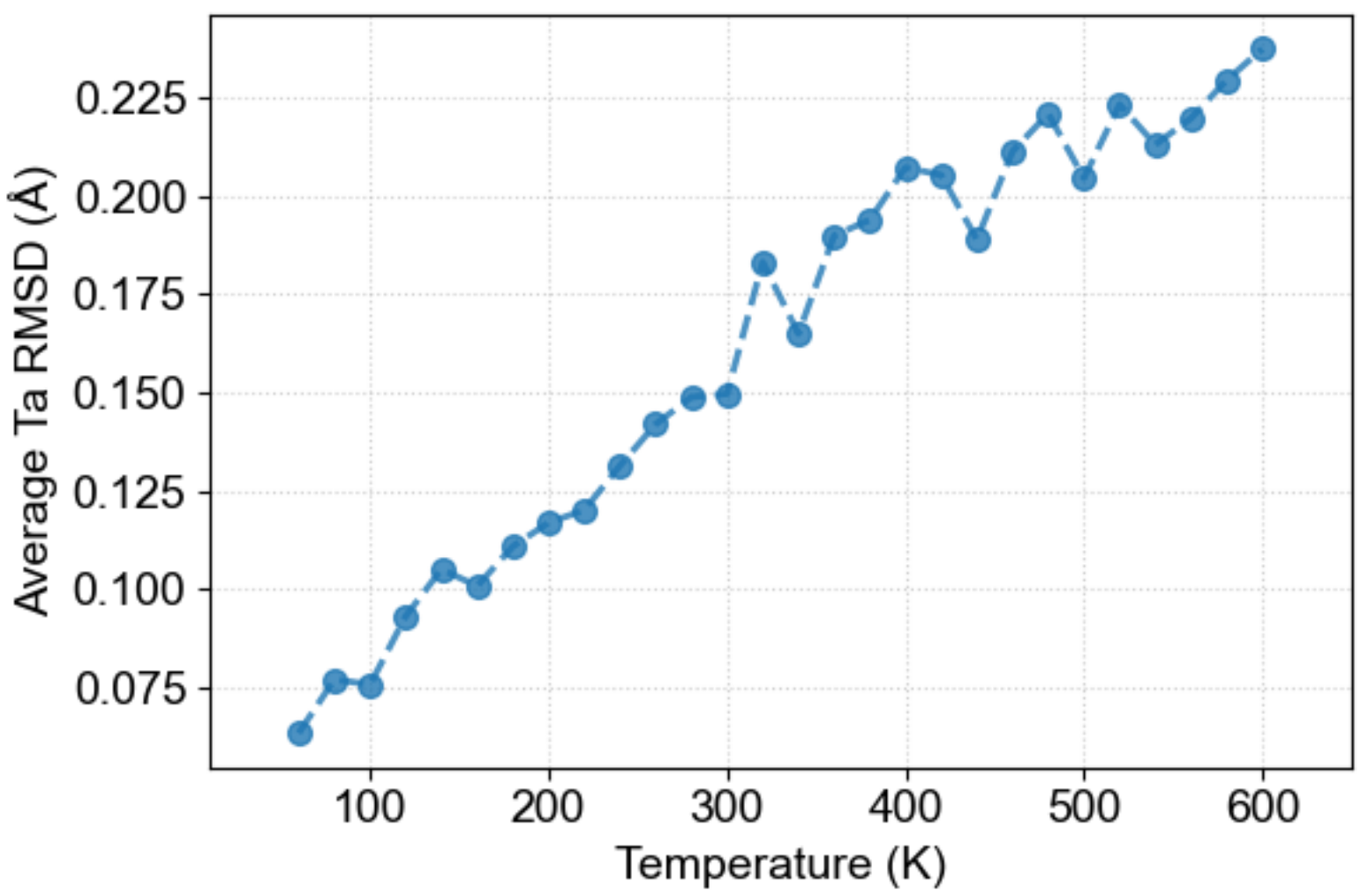

**Figure S4**. Average root mean squared displacement (RMSD) for Ta atoms in the last 2 ps of 15 ps of an MD simulation as a function of temperature.

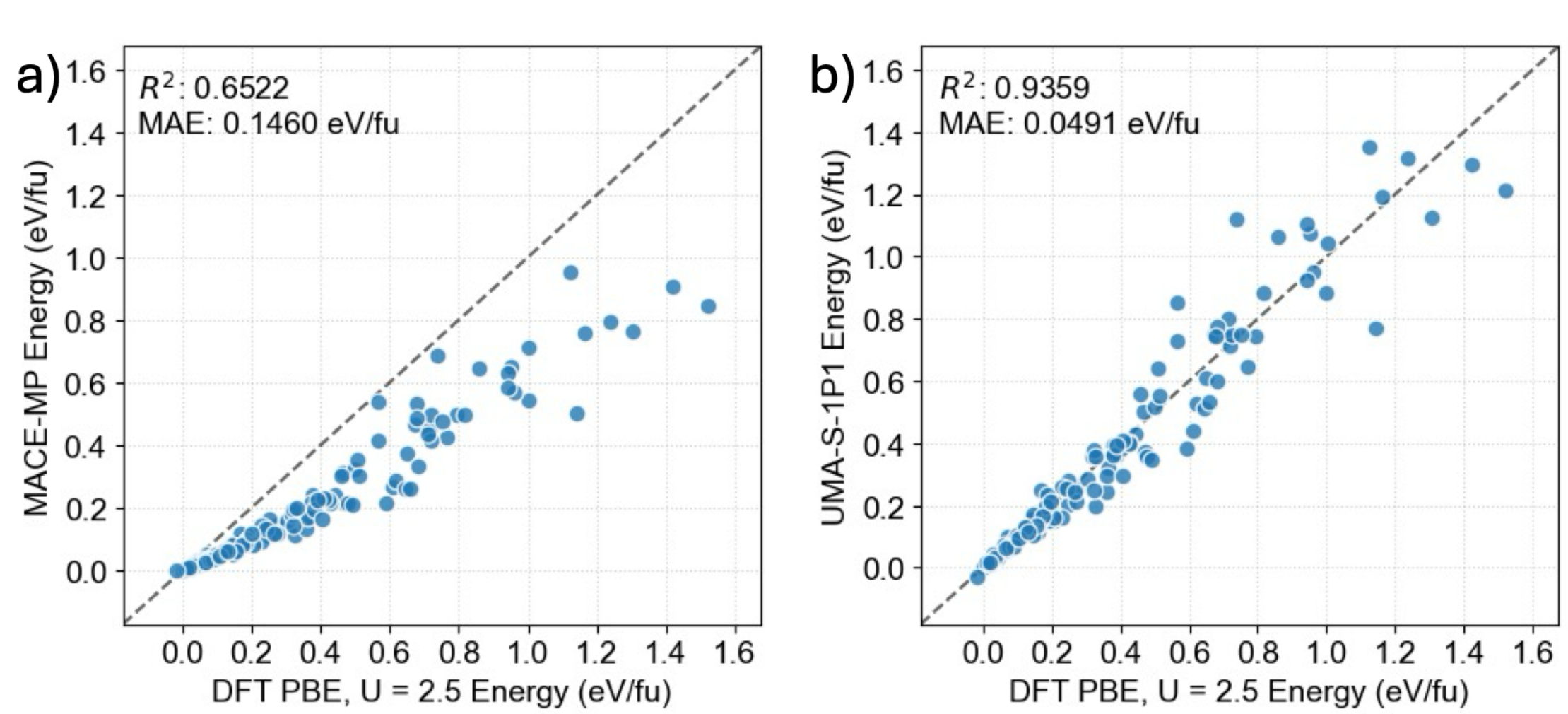

**Figure S5**. Benchmarking of (a) MACE-MP and (b) UMA s-1p1 models against spin-polarized DFT PBE with U = 2.5. The benchmarking dataset was generated by creating random displacements in both NM and CCDW 1*T*-$TaS_2$ structures in range from 0.01 Å to 0.15 Å.

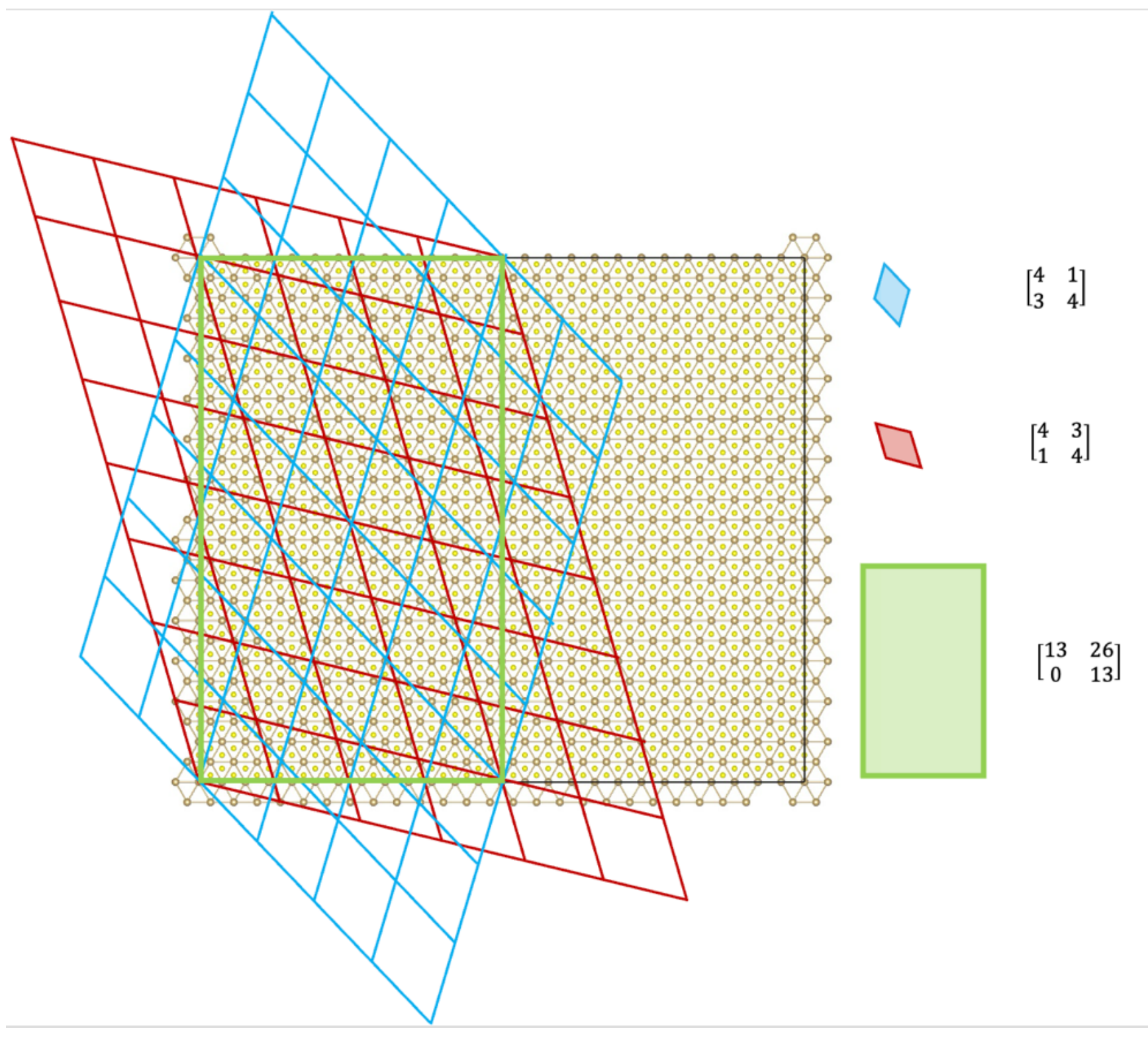

**Figure S6**. 1*T*-$TaS_2$ periodic cell accommodating both alpha (blue) and beta (red) CDW chiralities is shown in green. The final cell used in simulations is shown in the figure with the black outline and has the $\begin{bmatrix} 26 & 26 \\ 0 & 13 \end{bmatrix}$ transformation matrix.

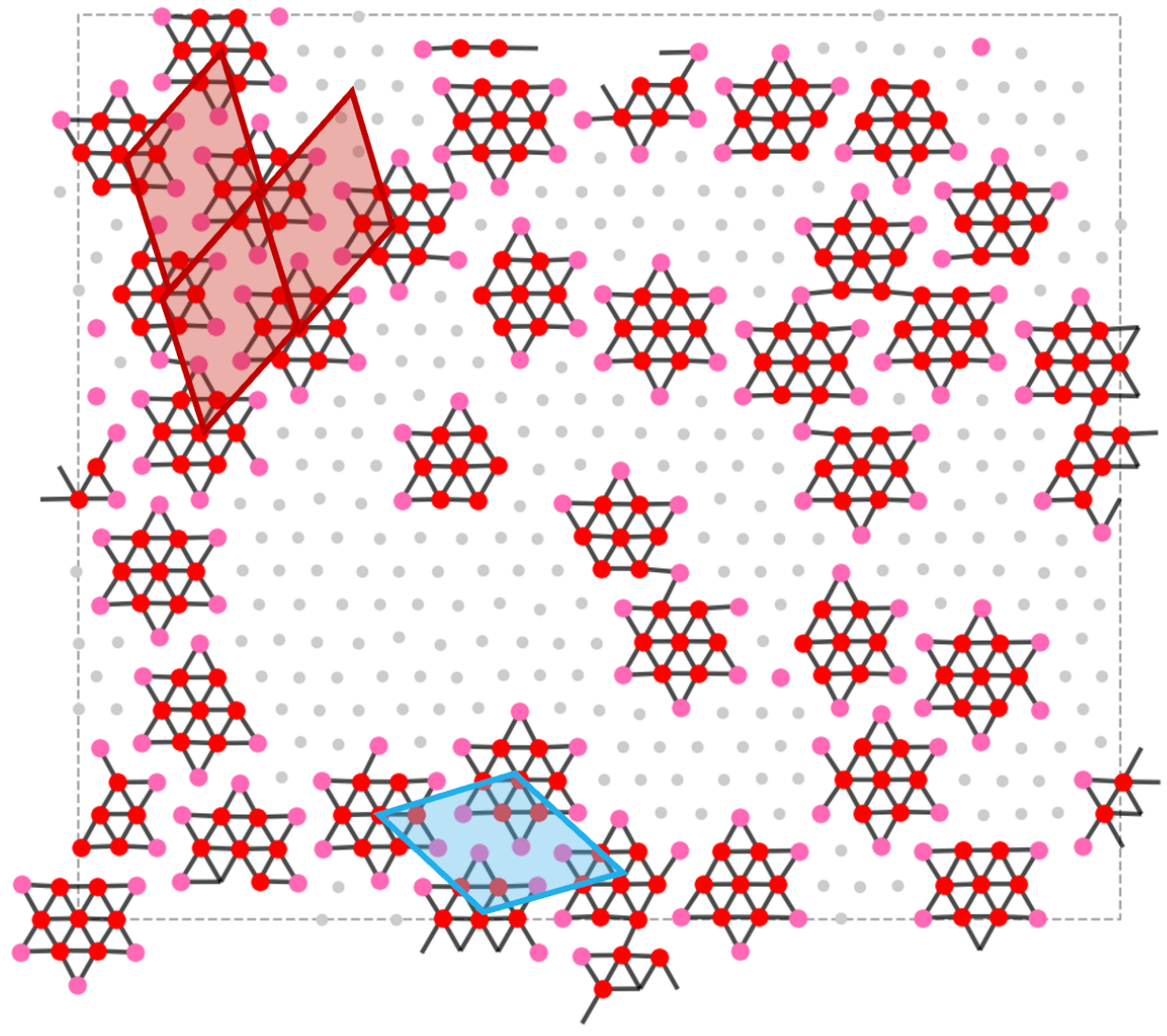

**Figure S7**. Snapshot of 1*T*-$TaS_2$ trajectory at T=140K (image is averaged across 5 ps), demonstrating alpha (red) and beta (blue) CDW domain chirality formation within the same simulation cell.

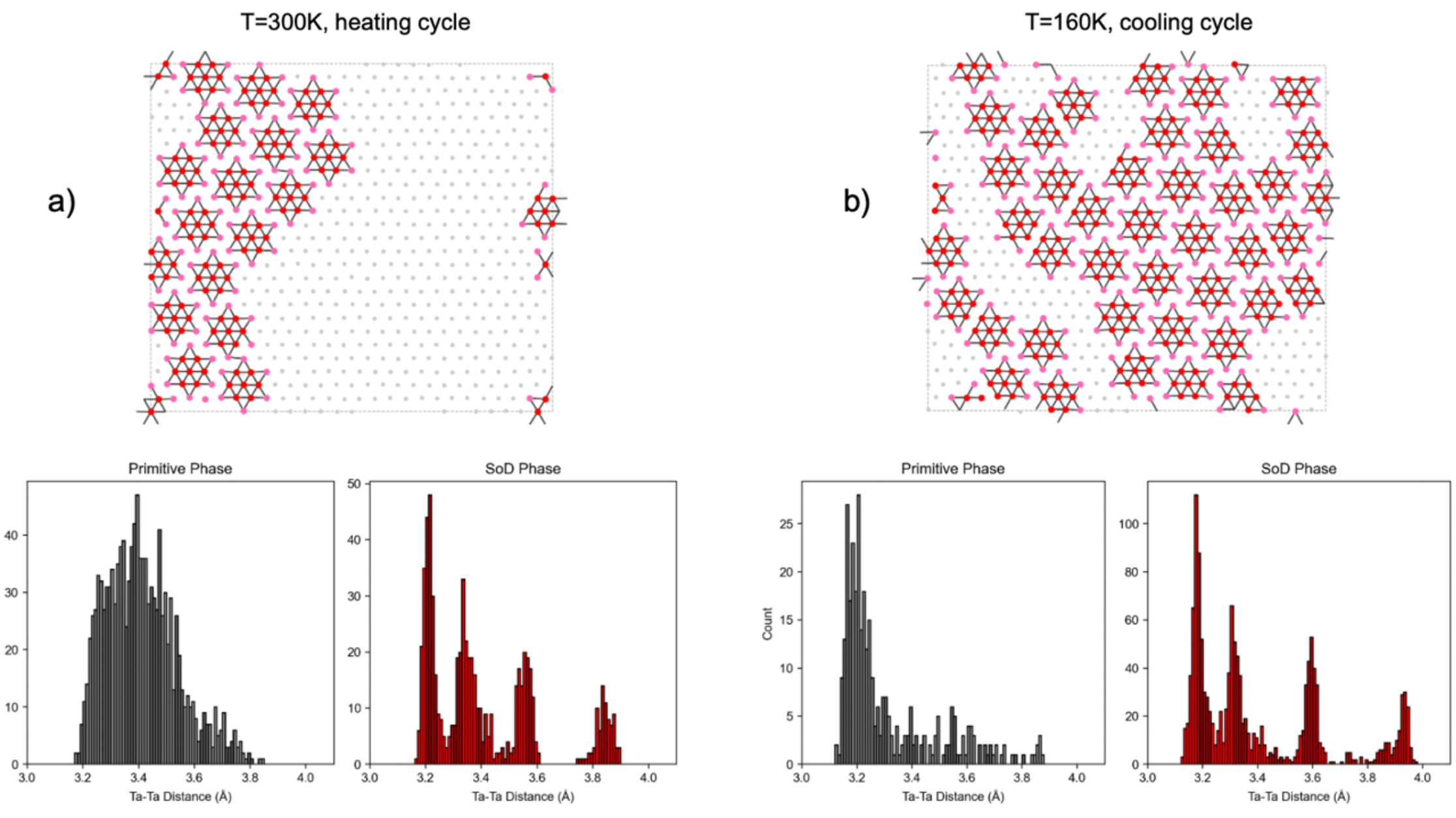

**Figure S8**. Ta atom positions averaged for 15 ps for (a) – T = 300 K and (b) – T = 160 K with Ta-Ta bond length distributions across different regions – CCDW-like and NM-like.

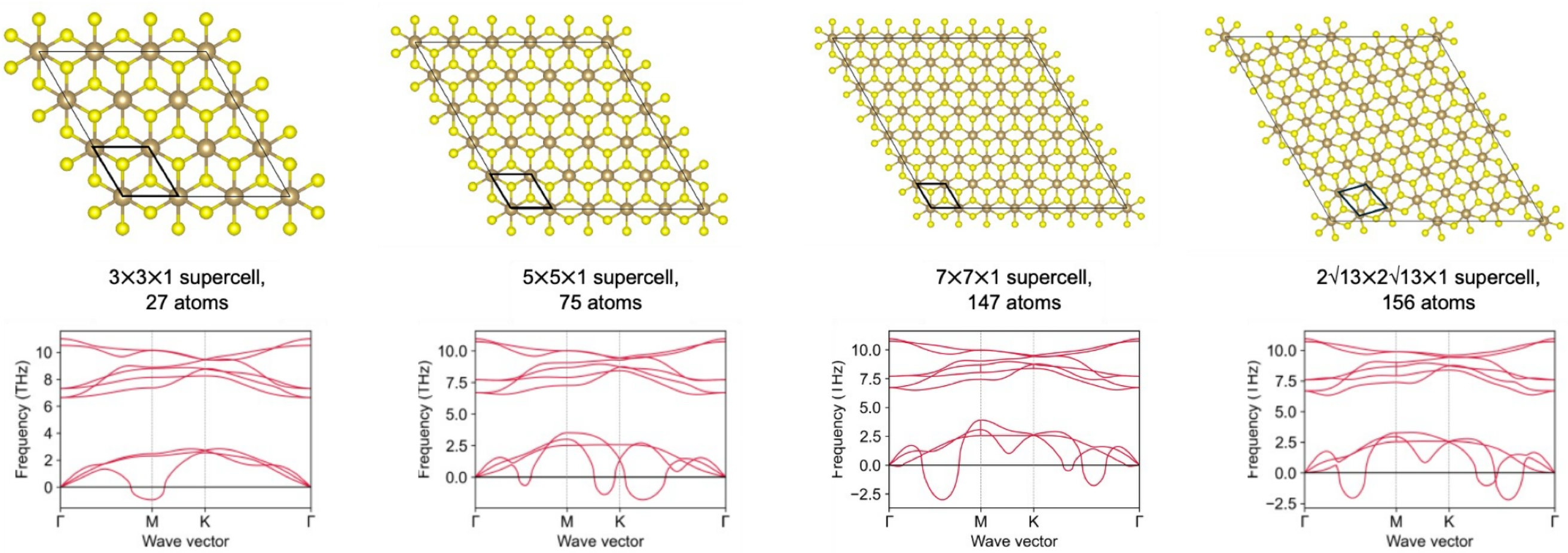

**Figure S9.** DFT-calculated phonon dispersions of the NM 1$T$-$TaS_2$ phase obtained from different supercells with the finite-difference method. Convergence is observed for the last two supercell sizes.

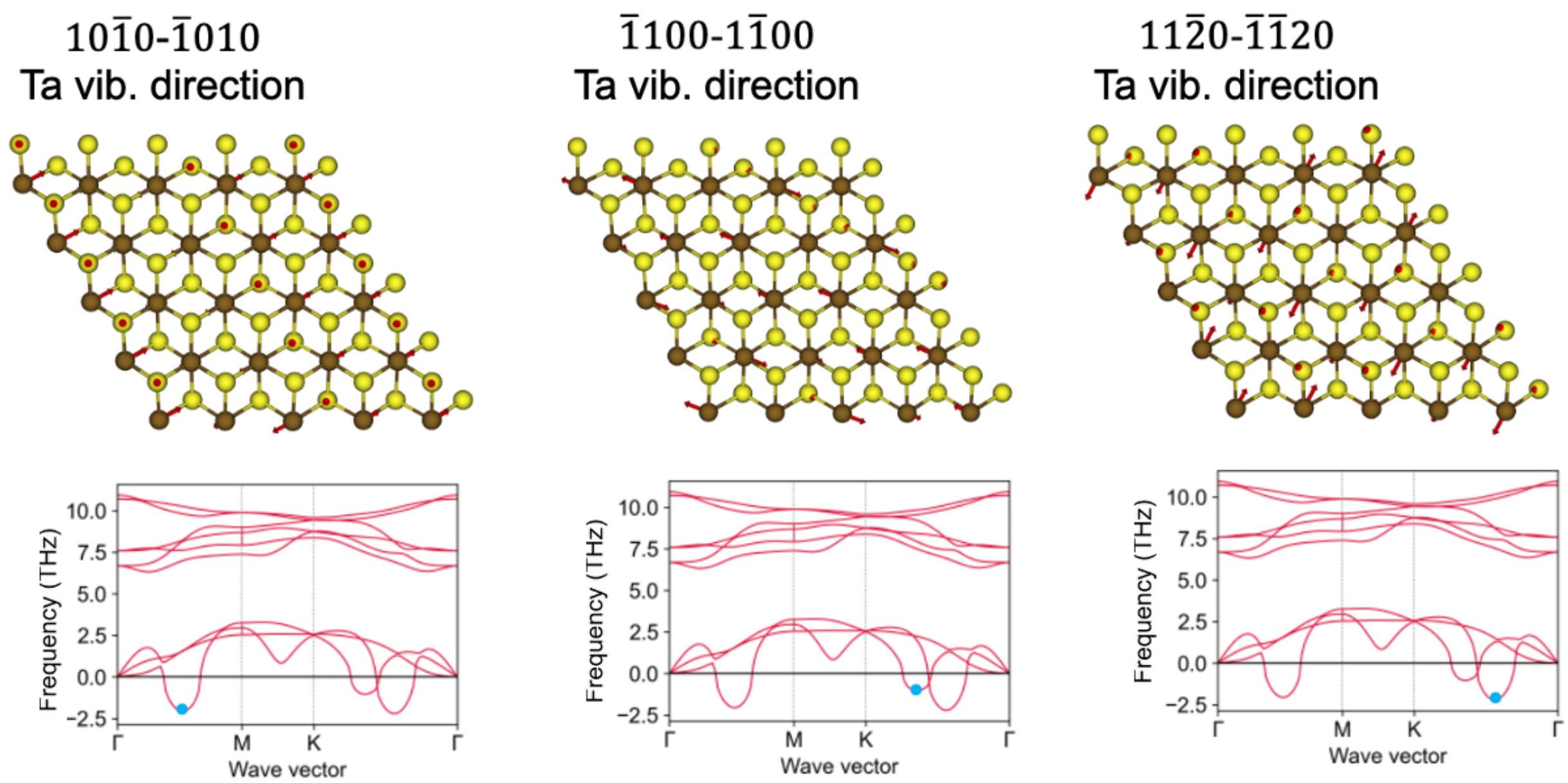

**Figure S10**. Different eigenvectors of soft modes of 1*T*-$TaS_2$ NM phonon spectrum.